\documentclass[aps,prl,twocolumn,showpacs,superscriptaddress,amssymb]{revtex4}

\usepackage{graphicx}
\usepackage{dcolumn}
\usepackage{textcomp}
\usepackage{amsmath}
\usepackage{hyperref}
\usepackage{natbib}
\usepackage{bm}

\hypersetup{
pdftitle    = {Control and coherence of the optical transition of single defect centers in diamond},
pdfauthor   = {Lucio Robledo},
}

\begin{document}

\title{Control and coherence of the optical transition of single defect centers in diamond}

\author{Lucio Robledo}
\email{l.m.robledoesparza@tudelft.nl}
\author{Hannes Bernien}
\author{Ilse van Weperen}
\author{Ronald Hanson}
\affiliation{Kavli Institute of Nanoscience Delft, Delft University of Technology, 
P.O. Box 5046, 2600 GA Delft, The Netherlands}

\date{\today}

\begin{abstract}
We demonstrate coherent control of the optical transition of single Nitrogen-Vacancy defect centers in diamond. On applying short resonant laser pulses, we observe optical Rabi oscillations with a half-period as short as 1 nanosecond, an order of magnitude shorter than the spontaneous emission time. By studying the decay of Rabi oscillations, we find that the decoherence is dominated by laser-induced spectral jumps. By using a low-power probe pulse as a detuning sensor and applying post-selection, we demonstrate that spectral diffusion can be overcome in this system to generate coherent photons.
\end{abstract}


\maketitle

Quantum control of light and matter is an outstanding challenge in modern science. Full control over both the spin state and the optical transition of a quantum system enables exciting applications such as spin-based quantum information processing and long-distance quantum teleportation of a spin state \cite{Olmschenk}. The basic operation underlying these applications, generation of non-local spin-spin entanglement, was recently demonstrated using individual trapped ions \cite{Monroe}. It is highly desirable to achieve the same level of control in solid-state systems as these may be easier to scale and promise higher integration density.

Candidate systems are required to have an electronic ground state with non-zero spin as well as a strong optical transition. Furthermore, coherent control over this transition is essential, as well as indistinguishability of the emitted photons \cite{HOM}. In this context, several promising systems are studied extensively. Two-photon interference from independent sources was demonstrated for donor-bound excitons in semiconductors \cite{Yamamoto} and for quantum dots \cite{Flagg}. Optical Rabi oscillations have been observed in quantum dots \cite{ROquantumdots} and Nitrogen-Vacancy (NV) color centers in diamond \cite{Batalov}. A major challenge for these solid-state systems is the inhomogeneous broadening of the optical transitions caused by fluctuations in the solid-state environment \cite{Hogele,Jelezko}. These fluctuations not only degrade the fidelity of optical control, but also reduce the visibility in two-photon interference experiments. Since this interference is crucial for generating non-local entanglement, imperfect contrast directly translates into a reduced entanglement fidelity.

Here, we demonstrate coherent optical control over the orbital state of single NV centers in diamond. Using short resonant laser pulses, we induce coherent oscillations of the optical transition. The decay of these oscillations yields important information on the transition's coherence \cite{DephasingMolecules}. We study the dependence of the coherence time on the involved laser fields. In particular, we find that spectral jumps, induced by an inevitable repump process, appear to degrade the fidelity of optical control. A scheme where the control is preceded by a detuning-sensitive stage is implemented. We show that post-selection based on the detuning sensing can be used to circumvent the degrading effect of spectral jumps on two-photon interference.

\begin{figure}
  \centering
  \includegraphics{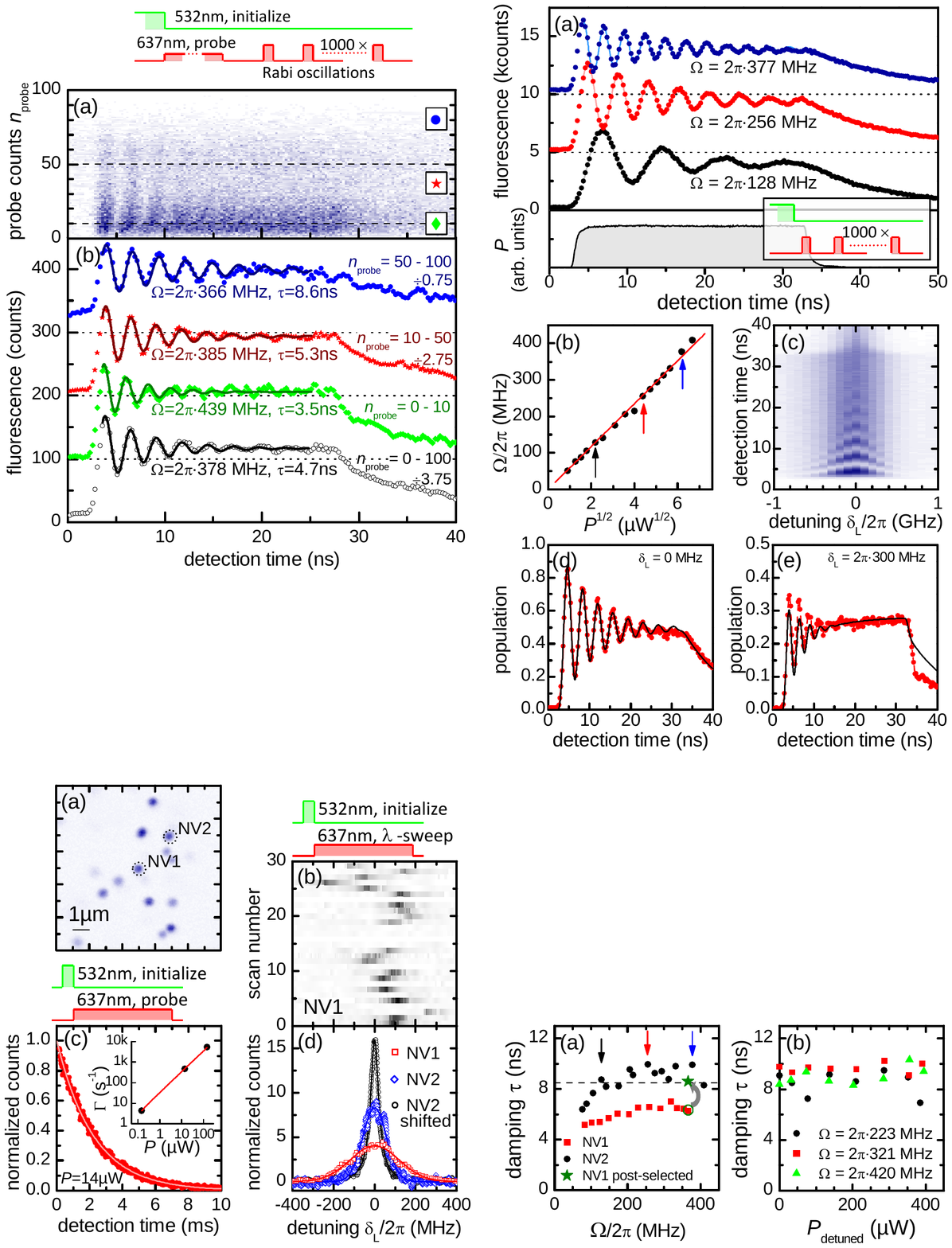}
  \caption{\label{fig:res_exc} (color online). {\bf(a)} Confocal fluorescence image of individual NV centers. {\bf(b)} Series of resonance scans. Green repump pulses applied in between scans to re-charge the NV center induce spectral jumps. {\bf(c)} Resonant excitation leads to exponential decay of fluorescence due to photo-ionization. Inset: Ionization rate $\Gamma$ is proportional to excitation power. {\bf(d)} Sum of individual excitation scans yields frequency distribution of spectral jumps. In comparison, for NV2 individual scans have been shifted to coincide in center frequency and then summed, to obtain the intrinsic linewidth. Data is obtained from an average of individual scans, normalized to the area under the curve.}
\end{figure}

The NV center consists of a single substitutional nitrogen atom in the diamond lattice, located next to a vacancy. We use a type IIa CVD-grown bulk diamond sample from Element Six ($\left\langle 100 \right\rangle$ oriented), containing a concentration low enough to detect individual negatively charged NV$^-$ centers (Fig.~\ref{fig:res_exc}(a)). 

The experiments are performed in a scanning confocal microscope using a flow cryostat at a temperature of $T=8$\,K. NV center fluorescence is excited either off-resonant using a $\lambda = 532$\,nm (green) laser or on resonance by means of a narrow-band tunable $\lambda = 637$\,nm (red) diode laser. For continuous resonant excitation, the laser frequency is actively stabilized using a high-resolution wavemeter. An electro-optical modulator (EOM) with risetime of 1.3\,ns allows us to apply short resonant pulses. The green laser is sent through an acousto-optical modulator with a risetime of 4\,ns. A dichroic mirror and additional longpass filters reject the excitation laser, so that only emission into the phonon sidebands (\,$>$\,660\,nm) is detected by an avalanche photodiode (APD) in the single-photon counting regime.

Excitation spectra are obtained by sweeping the excitation frequency across the NV center resonance and recording the red-shifted phonon-sideband emission. An excited state consisting of a spin triplet with sublevels ($S_{x,y,z}$) and orbital doublet with sublevels ($E_{x,y}$) gives rise to six resonances \cite{Manson}. However, in general strong spin-flip transitions lead to fast spin polarization, preventing most resonances to be directly observed in single-laser excitation experiments \cite{Tamarat}. Here, we show results from NV centers in the high strain regime ($\delta_{str} \approx 40$\,GHz \cite{footnote1}) where we excite the cycling ($S_z, E_x$) transition \cite{Tamarat}. This transition has a very low probability for spin-flips. After a spin-flip the same laser field can excite the ($S_x, E_x$) resonance detuned by $\approx$\,200\,MHz which suffers from strong spin polarization back into the $S_z$ state \cite{Manson}. In consequence, the ($S_z, E_x$) transition in good approximation behaves like a 2-level system \cite{Manson}.

Continuous resonant excitation quickly bleaches NV emission, which can be recovered by off-resonant green excitation. We measure the fluorescence decay rate as function of resonant laser power by applying a pulse sequence that alternates a $10$\,$\mu$s green pulse and 200\,ms resonant excitation. We find that the fluorescence during the resonant pulse decays exponentially at a rate proportional to the driving power, provided the excitation power saturates the NV transition (Fig.~\ref{fig:res_exc}(c)). Excitation at $\lambda = 637$\,nm but detuned from the exact resonance does not bleach the NV center. These results indicate that photo-ionization of the center out of the excited state is the origin of the observed decay. Off-resonant (green) excitation can excite charge impurities in the environment of the NV center (e.g. substitutional Nitrogen \cite{photoconductivity}) which can bring the NV center back to the negative charge state. This mechanism is intrinsic to the NV center, and explains the need for off-resonant repumping whenever driving NV centers resonantly. 

Fig.~\ref{fig:res_exc}(b) shows a series of resonant excitation scans. Between each scan, a green pulse was applied to prepare the NV center in the negative charge state and initialize the spin into the $S_z$ state \cite{Manson}. Large spectral jumps from scan to scan can be observed. In absence of the green pulse, spectral jumps remain below our experimental resolution (however the probability of ionization increases). This observation is consistent with the assumption that the green laser changes the local charge environment, leading to a shift of the resonance frequency via the DC Stark effect. The distribution of spectral jumps is thus dominated by the concentration of charge impurities in a center's direct environment. We observe large differences from center to center: In Fig.~\ref{fig:res_exc}(d), we compare the range of jumps for two centers. Spectral jumps of NV1 occur over a twice as large range as those of NV2, suggesting a higher concentration of nearby charge impurities. Summing individual scans while compensating for the spectral jumps reveals the Lorentzian-shaped absorption line in absence of green excitation. We find an intrinsic linewidth of 46$\pm$2\,MHz for both NV1 and NV2, presumably limited by interaction with the phonon bath at $T=8$\,K \cite{FuSantori}. 

In order to achieve coherent control, we apply resonant pulses. The sequence consists of a $10$\,$\mu$s green pulse to initialize the NV$^-$ charge and spin state, followed by 1000 repetitions of a 30\,ns resonant laser pulse and a 70\,ns off-time to allow the NV center to relax to its ground state before the next excitation cycle starts. We use a time-resolved single-photon counter to histogram the photon detection time relative to the start of the resonant laser pulse with a bin-size of \mbox{$\Delta t_{bin}=256$\,ps}. To build up statistics we sum over typically $10^9$ pulses. The number of counts during a time-bin is then proportional to the probability to occupy the excited state.

\begin{figure}
  \centering
  \includegraphics{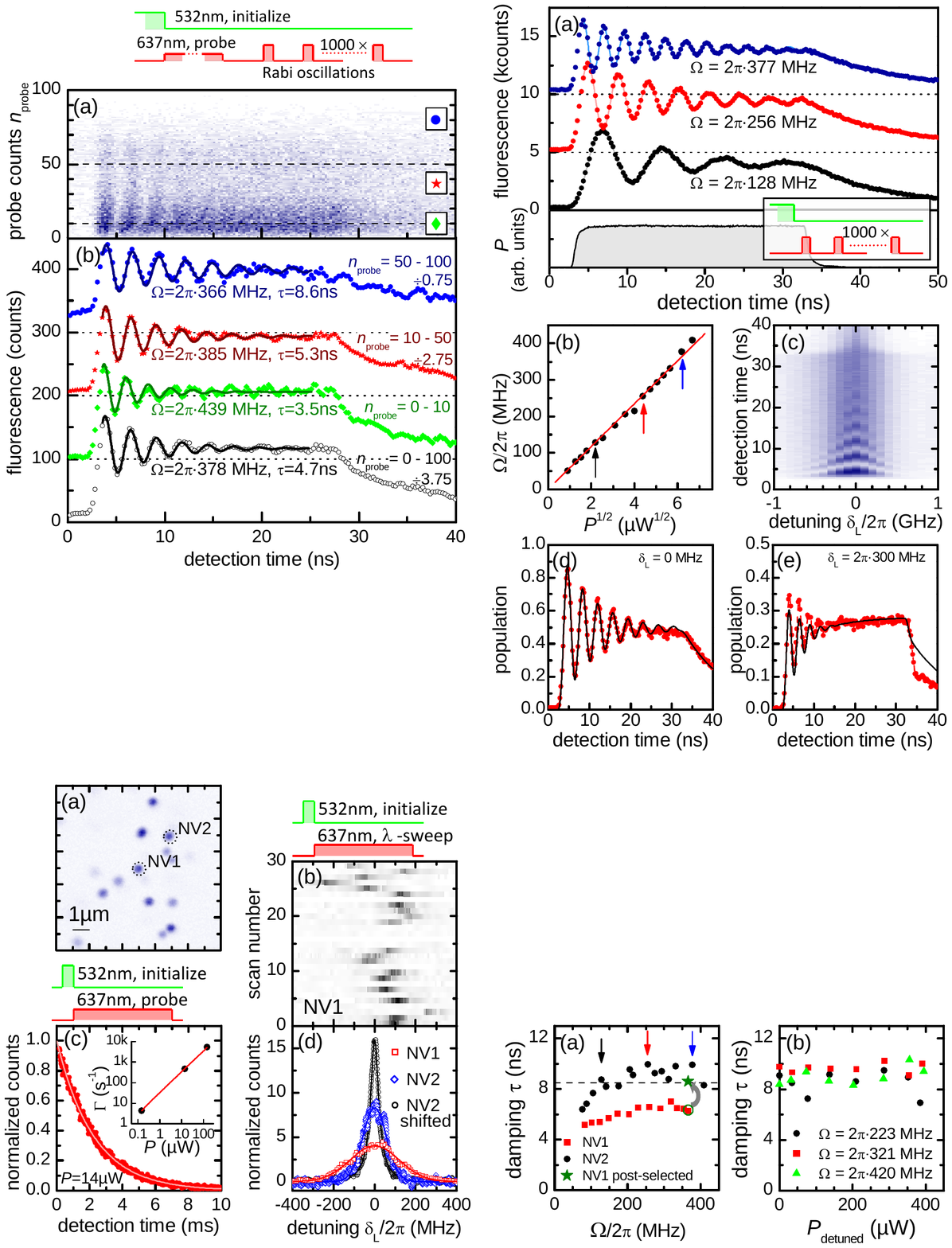}
  \caption{\label{fig:Rabioscillations} (color online). {\bf(a)} Rabi oscillations for \mbox{$P=38$\,$\mu$W,} \mbox{$P=19$\,$\mu$W} and \mbox{$P=4.7$\,$\mu$W.} The lowest dataset indicates the shape of the excitation pulse. {\bf(b)} Rabi frequency plotted versus square root of excitation power. {\bf(c)} Rabi oscillations as function of detuning for $\Omega_0=2\pi\cdot272$\,MHz. {\bf(d)}-{\bf(e)} Comparison of measured Rabi oscillations on resonance (d) and 300\,MHz detuned (e), and numerical data obtained from numerically integrating the optical Bloch equations.}
\end{figure}

During a resonant laser pulse, we observe coherent oscillations between the NV orbital states (Fig.~\ref{fig:Rabioscillations}(a)). On resonance, the oscillation frequency is given by \mbox{$\Omega_0 = \vec{\mu}\cdot\vec{E} / \hbar$}, with $\vec{\mu}$ denoting the NV zero-phonon-line dipole moment and $\vec{E}$ the electric field vector. The resulting square root dependence of oscillation frequency on excitation power is clearly observed (Fig.~\ref{fig:Rabioscillations}(b)). For the largest applied driving power, we can resolve $>$ 10 oscillations. The obtained Rabi frequency of $\Omega = 2 \pi \cdot 410$\,MHz corresponds to a $\pi$-pulse in only $t_{\pi}=1.2$\,ns. 

The oscillations are damped due to spontaneous emission and additional dephasing, such as phonon scattering. Experimental data fits excellent to an exponentially damped harmonic oscillation \mbox{$\cos(\Omega\cdot (t-t_0))\cdot$} \mbox{$\exp(-(t-t_0)/\tau)$}. On resonance, the exponential damping constant $\tau$ of the Rabi oscillations is determined by the pure dephasing time $T{_2}{^\ast}$ via $1/ \tau = 3/(4\,T_1) + 1/(2\,T{_2}{^\ast})$ \cite{SingleMoleculeSpectroscopy}. After turning off the driving field, the fluorescence decays exponentially due to spontaneous emission $\propto \exp(-t/T_1)$ with a lifetime $T_1$ of 10.9\,ns. 

In Fig.~\ref{fig:Rabioscillations}(c) we show Rabi oscillations for NV2 as function of laser detuning $\delta_L$. For detuned excitation the oscillation frequency increases to the generalized Rabi frequency $\Omega = \sqrt{{\Omega_0}^2 + {\delta_L}^2}$, while the oscillation amplitude decreases $\propto {\Omega_0}^2/{({\Omega_0}^2+{\delta_L}^2)}$ \cite{AllenEberly}. To accurately reproduce our experimental transients we numerically integrate the optical Bloch equations (Fig.~\ref{fig:Rabioscillations}(d),(e)) \cite{AllenEberly}, taking into account finite rise- and fall times of the driving field, and applying a spectral average over a Gaussian distribution of detuning values of FWHM $b=2\pi\,\cdot\,40$\,MHz. We assume a pure dephasing time of ${T_2}^\ast = 10$\,ns, consistent with the observed linewidth of $\Delta=46$\,MHz. The same simulation parameters are used to model the on-resonance as well as the 300\,MHz detuned transient. This simulation reproduces features for the detuned transients beyond the exponentially damped cosine, such as (i) the offset of the oscillations which increases with a negative exponential \cite{AllenEberly}, (ii) the pronounced kink arising from the finite fall-time of our excitation pulses and (iii) an increased damping rate. 

\begin{figure}
  \centering
  \includegraphics{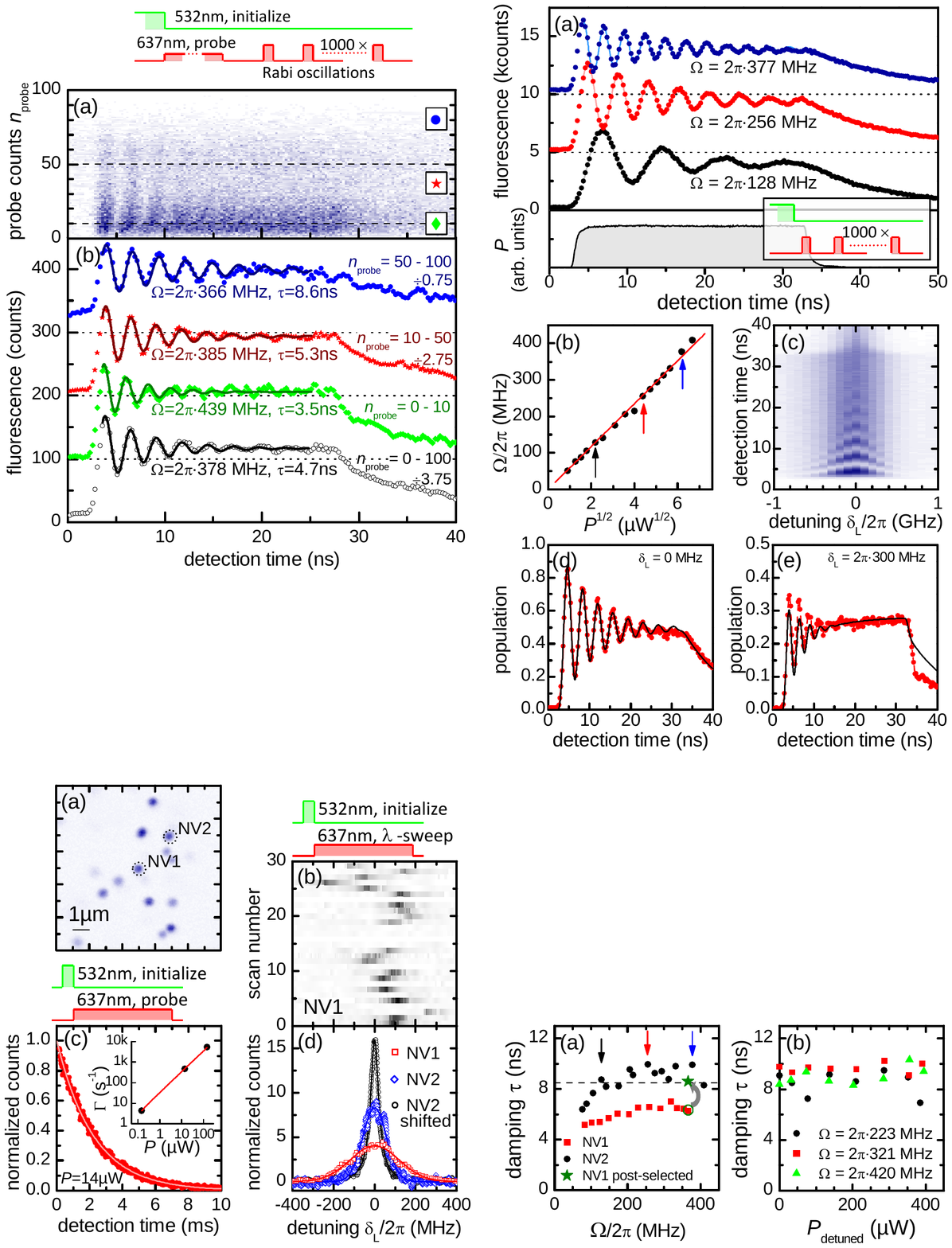}
  \caption{\label{fig:Coherence} (color online). {\bf(a)} Damping constant $\tau$ of Rabi oscillations as function of Rabi frequency. The dashed line indicates limit of $\tau$ for $T_2$ set by the single-scan linewidth $\Delta=48$\,MHz. Coherence is limited by spectral jumps (note difference between NV1 and NV2). Postselection can remove deteriorating effects of spectral jumps and enhance $\tau$ to the value expected from the linewidth. {\bf(b)} An additional laser field at $\lambda$=640\,nm is applied: This detuned laser has no influence on coherence.}
\end{figure}

In Fig.~\ref{fig:Coherence}(a) we show the observed damping constant of Rabi oscillations on NV1 and NV2 as function of the Rabi frequency. Coherence of Rabi oscillations is ultimately limited by fast dephasing processes on a nanosecond timescale, such as phonon scattering \cite{FuSantori}. However, for measurements that sum over several excitation cycles, such as the one in Fig.~\ref{fig:Rabioscillations}(a), we expect to be limited by slow processes such as the spectral jumps observed in Fig.~\ref{fig:res_exc}(b). The total measurement represents an average over different detuning values and thus different Rabi frequencies, which leads to a faster damping of the observed oscillations. Consequently, for NV2 we observe a better coherence than for NV1. For Rabi frequencies much larger than the spectral width of the distribution of resonances, the deteriorating effect of spectral jumps vanishes, since the Rabi frequency always largely exceeds the detuning. This is apparent in the initial increase of $\tau$ for NV2 which quickly settles to about 9\,ns for $\Omega > 2\pi\cdot 200$\,MHz. This value is close to the coherence expected from the linewidth (dashed line in Fig.~\ref{fig:Coherence}(a)). 

Previous research \cite{Batalov} suggested a decrease of coherence with increasing driving power. To investigate possible influence of the red laser on the coherence which is not related to the NV center resonance, we measured the damping of Rabi oscillations while also shining a detuned ($\lambda =640$\,nm) red laser onto the same NV center (Fig.~\ref{fig:Coherence}(b)). Even at an excitation power one order of magnitude larger than used for driving optical Rabi oscillations, we see no significant effect of the off-resonant laser. We conclude that for NV centers, optical coherence is not intrinsically limited by off-resonant effects of the driving field itself. 

\begin{figure}
  \centering
  \includegraphics{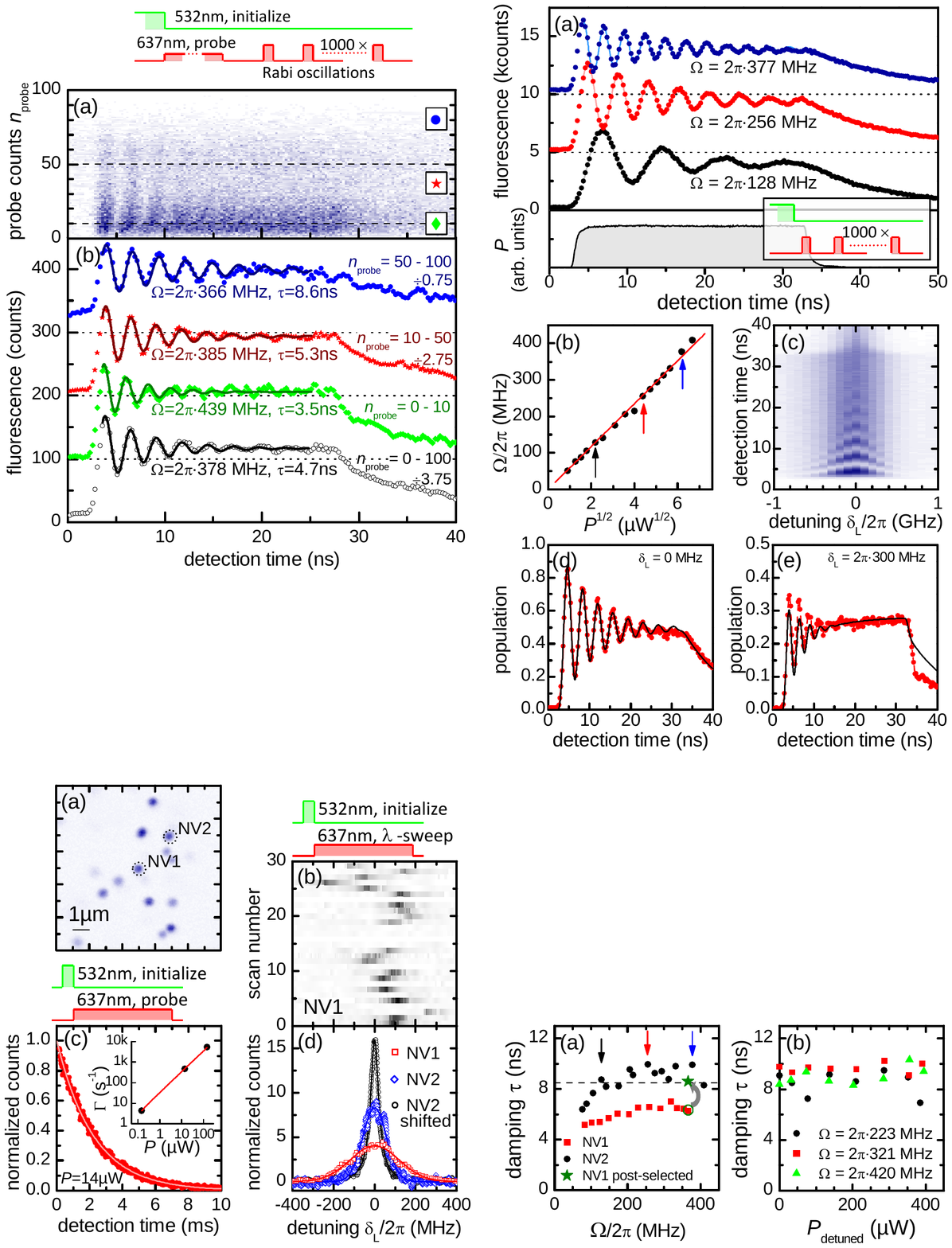}
  \caption{\label{fig:postselection} (color online). {\bf(a)} Detuning-resolved Rabi oscillations: detuning from resonance is derived from fluorescence counts during a 20\,ms weak resonant probe pulse. Subsequently measured Rabi oscillations are plotted as function of this count rate. Total data is obtained from $3 \cdot 10^7$ excitation cycles. {\bf(b)} Data from plot (a) is summed within three ranges. On resonance (high counts during probe pulse) oscillations show lower Rabi frequency and increased coherence. Transients have been scaled by the indicated factor for better comparison.}
\end{figure}

Summarizing these findings, the contrast of two-photon interference using NV centers will primarily be limited by frequency fluctuations of the emitted photons. We make use of the slow timescale of these fluctuations to show that even for NV centers suffering from unstable lines, the spectral stability (i.e. the coherence time) set by the single-scan linewidth can be reached. For that purpose we modify our measurement protocol: after each green initialization pulse we introduce an additional (weak) resonant laser pulse of t~=~20\,ms and count the photons $n_{probe}$ detected during that interval. These counts $n_{probe}$ now serve as a measure of the laser detuning during that specific interval. For large detuning we expect low $n_{probe}$, while resonance is identified by a large value of $n_{probe}$. We then record the time-resolved fluorescence during the strong (Rabi) pulses as a function of $n_{probe}$ (Fig.~\ref{fig:postselection}(a)). The measured data is divided into three regions of $n_{probe}$ and each of them is summed (Fig.~\ref{fig:postselection}(b)). For low $n_{probe}$ we identify the characteristic features of detuned excitation, as already observed in Fig.~\ref{fig:Rabioscillations}(e): increased Rabi frequency, increased damping, and a kink at the falling edge of the driving pulse. 
Post-selecting on-resonance events in turn leads to improved coherence: Already the damping time of the intermediate region of $n_{probe}$ outperforms the value for the overall data. Strikingly, selecting only the events with highest $n_{probe}$, the damping time is nearly doubled, and reaches the value expected from single-sweep linewidth measurements (Fig.~\ref{fig:Coherence}(a)). This scheme therefore allows for heralded two-photon interference for emitters with unstable resonances, provided that spectral jumps happen on a timescale which is slow compared to the photon detection rate. Compared to an interference experiment based on spectral filtering of the emission lines of both emitters to ensure resonance, this scheme yields higher interference contrast as detector dark counts only need to be accumulated while emitters are resonant to the driving field.

In summary, we demonstrated coherent control over the NV center's orbital state. Rabi frequencies \mbox{$\Omega>2\pi\cdot$400\,MHz} with more than 10 oscillations have been observed. Damping of Rabi oscillations is dominated by slow spectral jumps of the NV's resonance frequency, which however can be overcome by a detuning sensitive stage prior to coherent control. This result highlights the good prospects for quantum information processing based on probabilistic entanglement of distant NV centers using two-photon interference.

We thank V.V. Dobrovitski and M.D. Lukin for fruitful discussions. This work is supported by the Dutch Organization for Fundamental Research on Matter (FOM) and the Netherlands Organization for Scientific Research (NWO). L. R. acknowledges support of the European Community under a Marie-Curie IEF fellowship.


\begin{thebibliography}{}
\bibitem{footnote1} 
To determine the strain splitting, we measured the excitation spectrum of NV1 consisting of six resonances. Simultaneous MW excitation resonant with the ground state splitting of 2.88\,GHz was used to prevent spin polarization. All other measurements have been performed without additional MW excitation.
\end{thebibliography}
\end{document}